\documentstyle[prl,aps]{revtex}
\begin{document}
\draft
\preprint{\today}
\title{Sum rules and electrodynamics of high-$T_c$ cuprates in the
pseudogap state.}
            \author{ D.N.~Basov, E.J.~Singley, S.V.~Dordevic}
\address{Department of Physics, University of California,
San Diego, La   Jolla, CA 92093}

\wideabs{
\maketitle

\begin{abstract}
We explore  connections between the electronic density of states (DOS) in a
conducting system and the frequency dependence of the scattering  rate
$1/\tau(\omega)$ inferred from infrared spectroscopy. We show  that
changes in the DOS upon the development of energy gaps can be reliably
tracked through the examination of the $1/\tau(\omega)$ spectra using the
sum rules discussed in the text. Applying this  analysis to the charge
dynamics in high-$T_c$ cuprates we found radically  different trends in the
evolution of the DOS in the pseudogap state and in the superconducting
state. \end{abstract}
}

\narrowtext
One of the most enigmatic properties of cuprate high-temperature 
superconductors is the pseudogap  in the spectrum of low-energy excitations 
developing primarily in weakly doped materials at a temperature T* well 
above the superconducting transition $T_c$.\cite{statt} First discovered 
through experiments probing  spin-related behavior the pseudogap also leads 
to a characteristic modification of a variety of properties in the charge 
sector. Some aspects of the pseudogap have lead to an interpretation of 
this phenomenon in terms of a "precursor" of the superconducting gap 
whereas another school of thought is proposing scenarios   based on spin- 
or charge-density wave fluctuations.\cite{statt} Using the novel sum rule 
analysis of the optical constants of cuprates measured for {\bf E} vector 
along the CuO$_2$ planes we are able to put constraints on the possible 
microscopic origins of the pseudogap. 
                                         
It is generally agreed that the electromagnetic response of the
conducting CuO$_2$ planes is dominated by the nodal segments of the Fermi
surface which remain nearly unaffected by the pseudogap developing close to
the $(\pi;0)$ and $(0;\pi)$ regions. Nevertheless,  there are profound
consequences of the pseudogap state for the in-plane carrier dynamics in many cuprates. The spectroscopic signatures of the pseudogap are best resolved in the spectra of the frequency dependent scattering rate\cite{basov96,Puchkov}: 
 \begin{equation}
\frac{1}{\tau(\omega)}=\frac{\omega_{p}^{2}}{4 \pi}
Re\left(\frac{1}{\sigma(\omega)}\right),
\end{equation}
 where $\omega_p$ is the plasma frequency and 
$\sigma(\omega)=\sigma_1(\omega) +i\sigma_2(\omega)$ is the complex 
conductivity.\cite{ir} In the underdoped compounds (left panels in Fig.~1) 
the spectra of $1/\tau(\omega)$ are depressed at $T<T^*$ and reveal a 
characteristic threshold structure at $\Theta \simeq 500-600$ cm$^{-
1}$.\cite{basov96,puchkovprl96} At $\omega<\Theta$  the scattering rate 
measured at $T<T^*$ increases faster than linear with a crossover to a less 
rapid (nearly linear) increase at frequencies above the $\Theta$ threshold. 
This behavior is universal and is observed in a large variety of materials. 
\cite{basov96,Puchkov,puchkovprl96,tl2201,mcgire,startseva,singley01,shen-new} As doping progresses to optimal and overdoped similar depression of 
$1/\tau(\omega)$ occurs only at $T\ll T_c$(right panels in Fig.~1). 

Conventional interpretation of these results is based on phase space
arguments: a partial gap reduces the amount of final states available for
scattering of the nodal quasiparticles therefore leading to a depression
of $1/\tau(\omega)$ for energies below the pseudogap.  The above argument
is corroborated by a close correspondence between the spectra of the
scattering rate extracted from IR optics and the form of the electronic
self-energy obtained from the angular resolved photoemission spectroscopy
(ARPES).\cite{Norman} In this paper we propose a new analysis suggesting a
direct relationship between the behavior of $1/\tau(\omega)$ and the prominent
features of the electronic density of states (DOS) $D(\omega)$. To 
elaborate on a connection between the spectra of $1/\tau(\omega)$ (Eq.~1) 
and the essential characteristics of the DOS in the studied materials we 
first turn to the BCS model. We calculated the conductivity of an $s$-wave 
superconductor both in the clean and dirty limits\cite{laplae} and 
determined $1/\tau(\omega)$ from Eq.1. A striking result is that the form 
of $1/\tau(\omega)$ reproduces the key features of $D(\omega)$ in a BCS 
superconductor including the gap at $2\Delta$  followed by  a sharp peak 
(Fig.~2).  The divergence of the actual DOS at the gap edge is not found in 
$1/\tau(\omega)$.  Also, the peak in $1/\tau(\omega)$ may occur somewhat 
above the gap energy and its location depends on the strength of (impurity) 
scattering. Nevertheless, the overall character of the {\it single-particle 
density of states} is reproduced in the $1/\tau(\omega)$ spectra. Similar 
connection also holds for $d$-wave superconductors.\cite{dahm} 

An intriguing attribute of the model spectra displayed in Fig.~2 is
revealed by the integration of $1/\tau(\omega)$. It appears  that the
area removed from the intra-gap region is fully recovered due to the
overshoot of the spectra at $\omega>2\Delta$. The reason for the above
"sum rule" can be clarified by the following considerations. Expressing
$\sigma(\omega)$ Eq.~1 through the dielectric function
$\epsilon_1(\omega)+i\epsilon_2(\omega)$  one finds $ 1/\tau(\omega) =
[\omega_p^2/\omega][\epsilon_2/(\epsilon_1^2+\epsilon_2^2+1-2\epsilon_1)]$.
The term on the RHS in the latter expression is to be compared with the
integrand in the well-known sum rule\cite{mahan}:
\begin{equation}\int_0^\infty d\omega {{1\over{\omega}}
{{\epsilon_2}\over{\epsilon_1^2+\epsilon_2^2}}} = {\pi \over 2}
 \end{equation}
The difference between the two terms arising from  $1-2\epsilon_1$ in
the denominator of the $1/\tau(\omega)$ expression results in less than  1$\%$ correction at frequencies $\omega<0.5 \omega_p$. Therefore the balance between the areas associated with the intra-gap  region and the overshoot in the
$1/\tau(\omega)$ spectra  that can be expressed as:
\begin{equation}
\int d\omega[ 1/\tau^A(\omega)-1/\tau^B(\omega)] \simeq 0
 \end{equation}
is in fact expected from the sum rule arguments.\cite{SR,phonon} In Eq.~3
indexes $A$ and $B$ refer to different states of the studied system (e.g.:
normal, pseudogap, superconducting). If the ${\pi \over 2}$ sum rule (Eq.~2)
and Eq.3 are applied to the results plotted in Fig.~2 this analysis
provides an additional support for the idea of the direct correspondence
between the $1/\tau(\omega)$ spectra and the $D(\omega)$. Indeed, the BCS
model suggests that the states removed from the intra-gap region are
recovered at energies above the gap. This fundamental conservation is also
manifested in the behavior of $1/\tau(\omega)$ (Fig.~2). Moreover, in those
situations when $[1-2\epsilon_1] \ll [\epsilon_1^2+\epsilon_2^2]$ the DOS
conservation can be quantitatively verified using Eq.~3.\cite{mode}

In Fig.~3 we show an experimental example confirming connections between 
the structure seen in $1/\tau(\omega)$ spectra and the features of the DOS. 
We studied the response of the single crystals of Cr which is a spin-
density wave (SDW) antiferromagnet with the Neel temperature $T_N=312$ 
K.\cite{crspectra} The results of the scattering rate analysis are reported 
in Fig.~3 for the first time. We find that  at $T>T_N$ the absolute value 
of $1/\tau(\omega)$ increases as $\omega^2$ in accord with the Fermi-liquid 
(FL) theory. At 10 K the SDW gap is fully developed giving rise to a non-
trivial form of the $1/\tau(\omega)$ spectra. The scattering rate is 
suppressed at $\omega< 500$ cm$^{-1}$ but then overshoots the 320 K 
spectrum with a maximum at $\omega\simeq 900$ cm$^{-1}$. This behavior is 
similar to the results for the optimally doped cuprates  plotted in Fig.~1. 
The main difference is that cuprates show a linear "background" in 
$1/\tau(\omega)$ as opposed to the $\omega^2$ background seen in the data 
for Cr. It is appropriate to compare the overshoot in the $1/\tau(\omega)$ 
spectra with the calculations for a BCS superconductor since the BCS theory 
is believed to produce an accurate representation of the DOS in a SDW 
system. While unmistakable similarities are  revealed by such a comparison, 
it is important to keep in mind that in Cr only a part of the Fermi surface 
is affected by the SDW state.\cite{fawcet} Therefore the gap in the DOS is 
incomplete which may account for a more gradual increase of 
$1/\tau(\omega)$ in the vicinity  of the gap energy compared to theoretical 
prescriptions in Fig.~2. As pointed out above, the conservation of the  DOS 
may be reflected through the balance of the "intra-gap" and "overshoot" 
areas (Eq.~3) provided the $1-2\epsilon_1$ correction is insignificant. We 
found that in Cr the latter correction amounts to about 0.1 $\%$ of the 
$\epsilon_1^2+\epsilon_2^2$ at $\omega<10,000$ cm$^{-1}$. Thus it is hardly 
surprising that Eq.3 is fulfilled for Cr with the accuracy of 10$\%$ with 
integration limited to $\omega_c=1500$ cm$^{-1}$. Similar behavior of $1/\tau(\omega)$ spectra is also found in charge-density wave materials\cite{valla} as well as in the heavy fermion materials revealing hybridization gap.\cite{Dordevic} 

It is apparent from  Fig.2 and Fig.~3 that the spectra of $1/\tau(\omega)$  capture the gross characteristics of the density
of states in a conducting system especially  in those situations when the
DOS is (partially) gapped. We now return to the data for cuprates
focusing on the implications of Eqs.~2-3 for the  understanding of the
charge response across the phase diagram. As pointed out above, the
dominant features of the $1/\tau(\omega)$   spectra obtained for underdoped
crystals in the pseudogap state are similar to those seen at $T\ll T_c$  in
the optimally doped samples. On a closer examination one finds  differences
as well. Suppression of $1/\tau(\omega)$  at $\omega<\Theta$ in the
pseudogap state {\it does not} lead to the development of the overshoot
between the spectra  measured at $T\ll T^*$ and $T>T^*$. The data for the
optimally doped crystals taken at $T\ll T_c$ and $T\simeq T_c$ does reveal
an overshoot at $\omega \simeq 800$ cm$^{-1}$.\cite{overshoot} In the
latter case, the area under the overshoot is balanced out by the reduction
of $1/\tau(\omega)$ at low energies with the accuracy better than 10 $\%$
in  Tl$_2$Ba$_2$CuO$_6$ crystal and about 15 $\%$ in the 
YBa$_2$Cu$_3$O$_{6.95}$ sample. This  balance is expected from Eqs.~2-3
because the contribution of $1-2\epsilon_1$ term is about 0.1$\%$ of
$[\epsilon_1^2+\epsilon_2^2]$ for either of the above superconductors. Thus
the form of the $1/\tau(\omega)$ spectra for the optimally doped system is
in accord with the notion of the transfer of the states from the intra-gap
region to a peak at $\omega>\Theta$. This behavior can be naturally
attributed to the opening of the superconducting energy gap. Although, the
absolute values of the optical constants  for the underdoped compounds are
also in the regime when the Eq.~3 ought to be satisfied, data gives  no
indications even for a partial recovery of the area associated with the far
IR depression of $1/\tau(\omega)$ is the pseudogap state.

The fact that in cuprates the spectra taken above and below $T^*$
"violate" Eqs.~2-3 within the frequency range of Fig.~1    signifies  the
differences in the changes of the low energy DOS in the pseudogap state
from the changes associated with superconductivity. Essentially, the
analysis described above indicates that the states removed from $\omega<
\Theta$  at $T<T^*$ disappear from the energy interval reliably sampled in
our experiments (0.5 eV).\cite{scales} This result is in accord with the
tunneling studies of the temperature dependence of the electronic DOS in
underdoped cuprates.\cite{renner} Indeed, data obtained at $T\ll T_c$
reveals a transfer of the electronic states from the intragap region in the
quasiparticle peaks just above the gap. On the contrary, the formation of
the pseudogap only leads to a depletion of DOS without visible  traces of
recovery of $D(\omega)$ at least at $\hbar \omega <$ 0.1 eV. Specific heat
measurements are also in accord with our findings since these experiments
suggest a reduction of the DOS in the pseudogap state within the energy
window accessible to thermodynamic  probes( $\simeq 500$ K).\cite{loram}
Yet, another indication of an energy scale extending beyond 0.1-0.5 eV that is involved into the pseudogap
state response is provided by the IR studies of the interlayer $c$-axis
conductivity $\sigma_c(\omega)$. Non-conservation of the low-energy
spectral weight $N_{eff}(\omega) = \int_0^\omega d\omega'\sigma_c(\omega')$
for $\omega<0.2-0.4$ eV  is inferred from the oscillator strength sum
rule\cite{c-axis}; these $c$-axis  results parallel our findings deduced
from the examination of the in-plane scattering rate using Eqs.~2-3. Thus
one can conclude that both spectroscopic and thermodynamic studies of
cuprates reveal anomalous changes of $D(\omega)$  in the pseudogap state
distinct from the DOS effects associated with superconductivity. Hence,
these results   argue against a common origin of the pseudogap state and
superconducting state.

Further support for the distinct genesis of the pseudogap and of
superconducting gap is provided by the data for La$_{2-x}$Sr$_x$CuO$_4$
(La214) and Nd$_{2-x}$Ce$_x$CuO$_4$ (Nd214) materials. The scattering rate
analysis of the in-plane response of these systems reveals a characteristic
threshold structure at $\Theta \simeq$ 500-600 cm$^{-
1}$.\cite{startseva,singley01}  In the double- or triple-layered materials
the development of superconducting gap occurs approximately at the same
energy as the pseudo-gap structure. However, in La214 and Nd214 compounds
the  signatures attributable to superconductivity are confined
to energies below 40-50 cm$^{-1}$\cite{singley01,somal96} whereas the
pseudogap feature is essentially identical to what is seen in the double-
layered compounds with $\Theta \simeq 500-600$ cm$^{-1}$. It is difficult
to account for the difference by more than one order of magnitude  between
the energy scales associated with the pseudogap and with superconductivity
if both effects result from the same microscopics. The
formation of the pseudogap is commonly discussed in the context of the
density wave ideas. Our analysis in conjunction with the data
for a  canonical SDW system such as Cr argues against this point of view as
well. Data shown in Fig.~3 is a direct testimony to a conventional transfer
of the intragap states to the energy region right above the gap. Similar
overshoot is also found in the quasiparticle relaxation of a charge-density
wave system.\cite{valla} Therefore, both spin and charge density waves are
unable to account for the experimental situation in the pseudogap state in
high-$T_c$ cuprates.

In conclusion,  the sum rule analysis of the optical constants based on 
Eqs.2-3  argues  against the decisive role of  both superconducting and 
density wave fluctuations in the formation of the pseudogap. We emphasize 
that the scale associated with the spectroscopic signatures of the 
pseudogap dramatically exceeds the characteristic temperature $T^*$. Such a 
disproportion between $\omega$- and $T$-scales may signal relevance of the 
many-body effects to the formation of the pseudogap state.\cite{QHE} 
Authors are grateful A.V.Chubukov, J.P.Carbotte, D.Pines,  T.Timusk, and 
especially to D.van der Marel for useful discussions. This work 
is supported by the DoE.

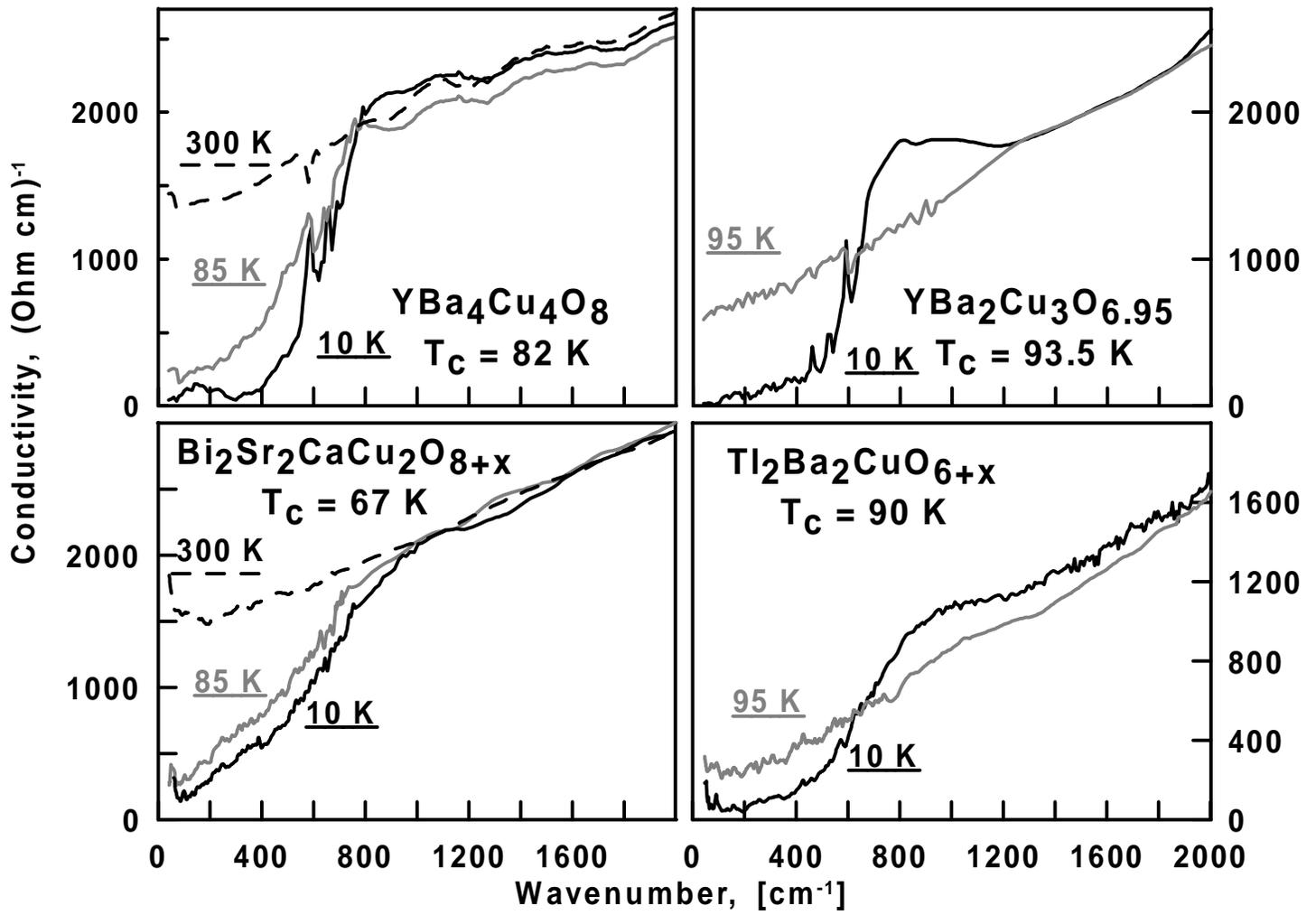
\begin{figure}
\caption{Spectra of the scattering rate determined for a variety of
cuprates. Left panel displays data for underdoped materials:
YBa$_2$Cu$_4$O$_8$[2] and Bi$_2$Sr$_2$CaCu$_2$O$_{8+x}$[5]. Right panel
shows data for the optimally doped compounds: YBa$_2$Cu$_3$O$_{6.95}$[2]
and Tl$_2$Ba$_2$CuO$_{6+x}$[6]. Solid lines: $T=10$ K, gray lines: $T\simeq
T_c$, dashed lines: $T=300$ K. The in-plane response of underdoped
materials in the pseudogap state is characterized by a depression of
$1/\tau(\omega)$ at $\omega<500$ cm$^{-1}$. In the optimally doped
materials similar depression is observed only in the superconducting state.}
 \end{figure}

 \begin{figure} \caption{The frequency dependence of the scattering rate
(Eq.~1) for a BCS superconductor in clean and dirty limit (black lines). In
the normal state (gray lines) we assumed $1/\tau(\omega)=const$ in accord
with the Drude model. Area balance in $1/\tau(\omega)$ spectra holds irrespective of the absolute values of the scattering rate. This attribute of the model spectra is exemplified in the right bottom panel displaying integrals of $1/\tau(\omega)$ at $T=T_c$ and $T=0$ as a function of the cut-off frequency.} \end{figure}

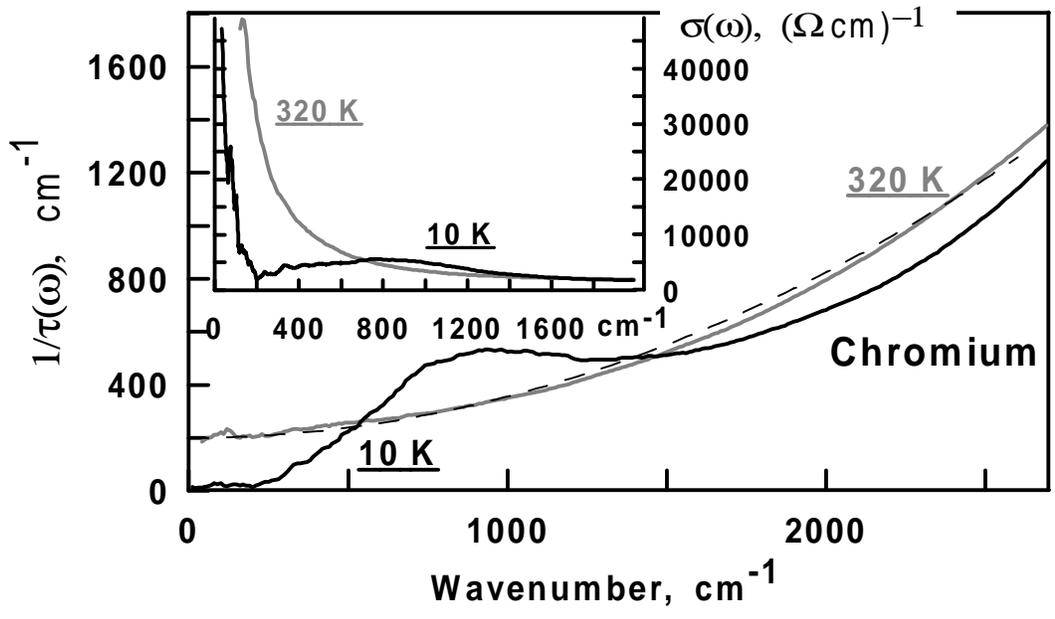
\begin{figure}
\caption{Top panel: $1/\tau(\omega)$ spectra for Cr. At $T>T_N$ data
follows the form $1/\tau(\omega) \propto \omega^2$ (dashed line) in
agreement with the Fermi liquid theory. At $T=10$ the SDW gap is fully
developed leading to non-trivial form of the $1/\tau(\omega)$ spectrum
described in the text. Inset shows the conductivity spectra. }
\end{figure}

\begin{references}
\bibitem{ir} The conductivity spectra are usually obtained from Kramers-
Kronig (KK) analysis of the raw reflectance data. Extrapolations to high
and low energies required for the KK transformations have only a minor
impact on the $1/\tau(\omega)$ spectra in the frequency region where the
actual data exist. For a review see for example Ref.[3].
\bibitem{SR} The scattering rate defined through Eq.~1 and the effective
mass $m^*=(\omega_p^2/\omega)(Im(1/\sigma))$ constitute a pair of optical
constants following Kramers-Kronig relations and provide a complete
description of electromagnetic response of a solid. Numerous sum rules for
other optical constants ($\sigma(\omega)$, $\epsilon(\omega)$, etc) all can
be viewed as consequences of the causality of the response that is
formalized through the KK integrals. Therefore a sum rule for
$1/\tau(\omega)$ is not an unexpected outcome of basic considerations of
the dispersion relations.
\bibitem{phonon} Electron-boson interaction can lead to a nearly parallel 
off-set of the $1/\tau(\omega)$ spectra at different temperatures that is 
not accompanied by noticeable modifications of the $\omega$-dependence 
(Fig.5 In Ref.[3]). This example suggests that data analysis based on  
Eq.~2-3 has to be practiced with caution. We believe that the application 
of Eq.~2-3 produces practically useful  results primarily in those cases 
when the electronic DOS undergoes radical changes such as an opening of the 
(partial) energy gap.
\bibitem{scales} This energy scale exceeds the antiferromagnetic exchange 
energy $J$ (typically 0.15 eV) and is of the order of the energies of the 
interband transitions in cuprates.
\end{references}
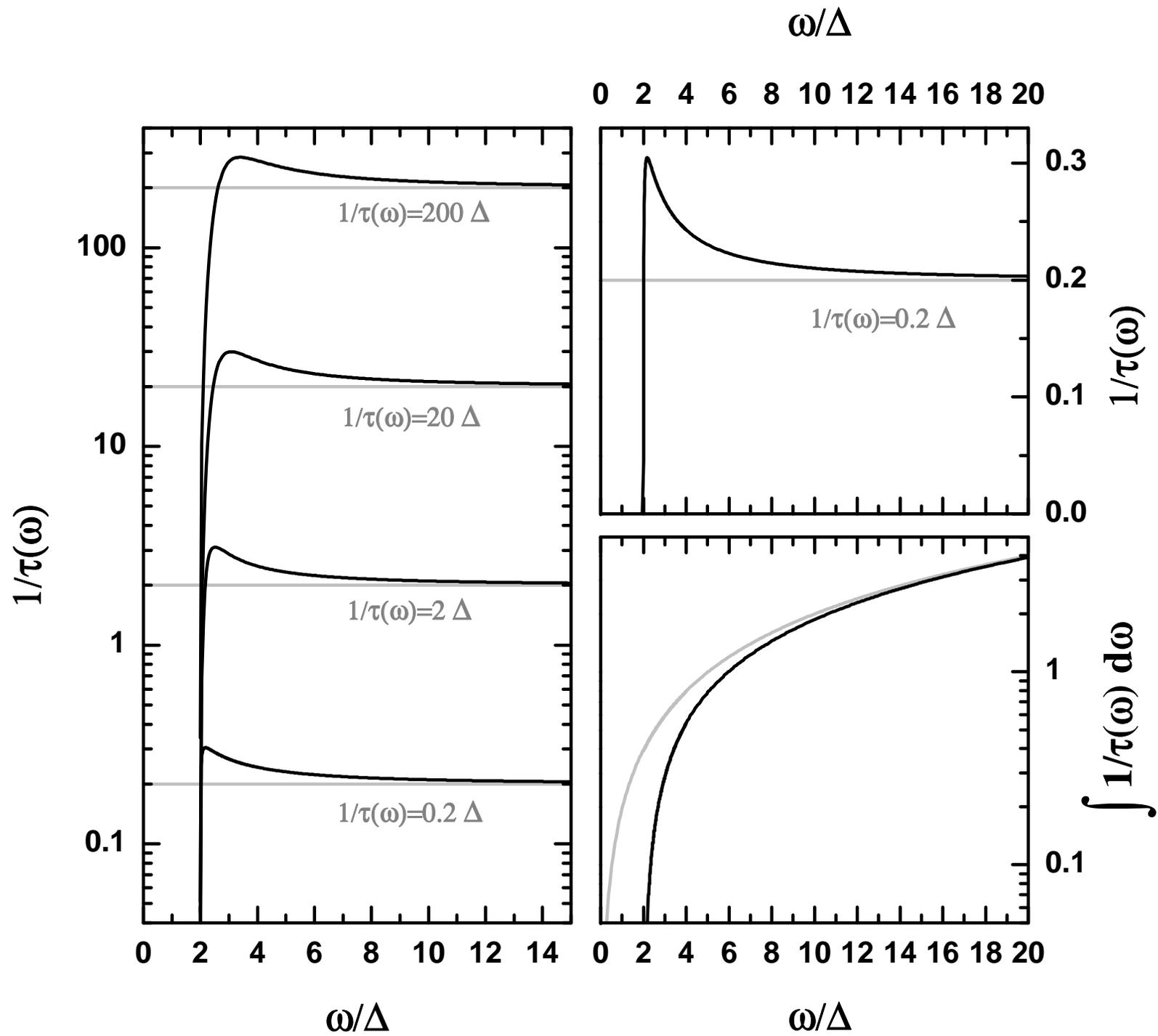
\end{document}